\documentclass{emulateapj}

\usepackage{pslatex}
\singlespace
\usepackage{natbib}
\usepackage{amsmath}
\usepackage{graphicx}
\begin{document}
\title{Radio Astrometry Of The Close Active Binary HR5110}
\author{$^1$E. Abbuhl, $^1$R. L. Mutel, $^1$C. Lynch,  $^2$M. G\"uedel}
\affil{$^1$Department of Physics and Astronomy, University of Iowa, Van Allen Hall, Iowa City, Iowa 52242, USA}
\affil{$^2$Department of Astronomy, University of Vienna, Vienna, Austria}

\begin{abstract}
The close active binary  HR 5110 was observed at six epochs over 26 days using a global VLBI array at 15.4~GHz. We used phase-referencing to determine the position of the radio centroid at each epoch with an uncertainty significantly smaller than the component separation. After correcting for proper motion and parallax, we find that the centroid locations of all six epochs have barycenter separations consistent with an emission source located on the KIV  secondary, and not in an interaction region between the stars or on the F primary.  We used a homogeneous power-law gyrosynchrotron emission model to reproduce the observed flux densities and fractional circular polarization.  The resulting ranges of mean magnetic field strength and relativistic electron densities are of order 10 G and $10^5$ cm$^{-3}$ respectively in the source region.

\end{abstract}
   
\section{Introduction}

Late-type stars in short-period binary systems often exhibit  signposts of enhanced magnetic activity. These include persistent photospheric spots, ultraviolet chromospheric emission, x-ray emission from hot coronae, and non-thermal radio emission \citep[e.g.][]{Guinan:1993}. The conventional interpretation is that tidal effects enforce spin-orbit coupling, and the consequent rapid rotation  drives a robust magnetic dynamo on one or both components of the system \citep[e.g.][]{Schrijver:1991}. 

Close active binaries are categorized by whether one of both companions fill their respective Roche lobes.  The eponymous RS~CVn systems have both components detached and no mass transfer \citep{Hall:1978}, while Algol systems are semi-detached with one component (typically the cooler, more evolved star) filling its Roche lobe, leading to episodic mass transfer.  Finally, contact binaries have both components within each other's Roche lobes, and continuously transfer mass. Curiously, radio luminosities do not differ greatly among these three classes \citep{Umana:1998}, while the mean X-ray luminosity of Algol-class systems is 3-4 times lower than RS CVn's \citep{Singh:1996}. This suggests that dynamical processes such as mass transfer and accretion disks may be decoupled from emission processes related to enhanced magnetic fields.

An important open question for these systems is whether the magnetically-driven enhanced emission arises primarily from a single active component or from a magnetic interaction region between the stars. Very-long-baseline radio interferometry (VLBI) provides a powerful method to address this question, since for nearby ($d\lesssim$ 100 pc)  systems, it can probe the  structure of the radio coronae at spatial scales smaller than the binary orbit.  A number of close active binaries have been studied using VLBI arrays for several decades \citep[e.g.,][]{Clark:1976, Mutel:1984,Lestrade:1984,Mutel:1985, Lestrade:1988, Massi:1988,Trigilio:1993,Lebach:1999,Ransom:2003}. However, locating the radio structure within the binary system requires high-precision astrometric techniques such as rapidly switching between the target star and one or more extragalactic phase reference sources with precisely determined positions.  Multi-epoch observations over several years are needed in order to simultaneously solve for the astrometric parameters (parallax, proper motion, orbital motion) with sub-mas accuracy.

The first measurements of this type were made by \citet{Lestrade:1993}, who showed that the radio centroid of  Algol 's inner binary moved with the KIV secondary. Subsequent astrometric VLBI studies of Algol \citep{Peterson:2010, Peterson:2011} confirmed the K star association. They also  showed that during flares,  the radio structure consisted of a large coronal loop approximately one stellar diameter height, with a base straddling the sub giant and oriented toward the B star. Other recent astrometric VLBI studies of close active binaries include UX Arietis \citep{Peterson:2011}, and IM Peg \citep{Ransom:2012}. In all three cases, the radio emission  is associated with the more evolved star,  a K subgiant or giant.

In this paper we report on an astrometric VLBI study of a fourth close active binary.  HR 5110 (BH CVn, HD118216) is a short period (2.61~day) binary composed of a F2 IV  primary and K2 IV secondary.  The orbit is circular and is oriented nearly face-on to the observer \citep{Eker:1987}. This orientation is well-suited to investigate whether the radio emission is associated with a single component or with an interaction region. It also provides a convenient test of active region association with polar spots \citep{Huenemoerder:2009}, assuming the magnetic and rotational axes are parallel and normal to the orbital plane, and hence oriented toward the observer.

Although no spots have been directly detected on HR5110, \citet{Little-Marenin:1986} found that the IR spectrum can be best-fit by assuming a persistent spot on the K secondary with a filling factor $\sim$0.25.   It has  been categorized  both as an RS CVn system \citep{Hall:1978} and  an Algol system \citep{Little-Marenin:1986}, although UV observations have found no direct evidence of a  accretion flow or disk. However, \citet{Little-Marenin:1986} argue that the secondary must fill its Roche lobe based on the spectral classification and its close binary separation.

There have been several estimates of the size and location of the active region in HR5110.  Previous VLBI studies of the radio corona \citep{Mutel:1985,Ransom:2003} found a `core-halo' structure, with a compact core size of order the K-star radius and an more extended halo somewhat larger than the orbital separation. This is similar to the results of \citet{Umana:1993}, who measured the radio spectrum of HR5110 between 1.4~GHz and 15~GHz during several epochs of moderate activity. They found that the spectral energy distribution could be fit with a two component gyro-synchrotron model  consisting of higher density core with a size about one-half the K-star radius and a lower density extended component approximately the size of the binary system.   

Observations at other wavelengths provide additional constraints on the active region's size and location.  \cite{Little-Marenin:1986} found that UV emission lines are best-fit with the K secondary's chromosphere and transition region, hence locating UV flares on the K star.  \citet{Mullan:2006}  modeled UV flare observations with a compact active loop whose length was smaller than 0.42 stellar radii. However, \citet{Graffagnino:1995}  argued that the active region may be much larger. They fit ROSAT X-ray data from a large outburst lasting more than three days. They found that a very large emission region, on the scale of the binary separation, is required to fit the observed light curve. They suggest that the flare may have been driven by magnetic reconnection in the interbinary region between the stars. Assuming the radio, UV, and X-ray emission regions share a common active-region site, an open question is: Where in the binary system does this region lie?

We observed  HR5110 at multiple epochs using an intercontinental phase-referenced VLBI array.  The observations were made at a frequency of 15~GHz, which provided twice the angular resolution of any previous VLBI study of this system. The primary goal of the observations was to determine the location of the radio emission within the binary system. We also modeled the observed emission properties (flux density, circular polarization, angular size) using a power-law gyrosynchrotron emission model to derive constraints on  plasma parameters in the source region.

\section{Observations and Data Analysis}

HR5110 was observed at six epochs of ten hours each with a global VLBI array consisting of the ten 25 m telescopes of the Very Long Baseline Array (VLBA)\footnote{The National Radio Astronomy Observatory is a facility of the National Science Foundation operated under cooperative agreement by Associated Universities, Inc.} and the 100 m Effelsberg telescope near Bonn, Germany.  We scheduled observations with a cadence that allowed sampling at approximately evenly spaced orbital phases.   The receivers sampled 64$\times$2 MHz channels using 2-bit sampling centered at 15.4 GHz in dual polarization mode. The synthesized beam was 0.4 mas in its smallest dimension which is approximately one fifth the binary's major axis (Table \ref{table:hr5110-params}). 

\begin{deluxetable*}{l c c l}[h]
\tablecaption{HR5110 properties and orbital elements}
%\tablewidth{5.in}
\tablehead{  \colhead{Parameter} & \colhead{Symbol} & \colhead {Value}  & \colhead{Ref.\tablenotemark{a}}  }
\startdata
% Insert table output here
\cutinhead{Fixed values}
	               Parallax & $\Pi$ & 22.21 mas & 1 \\ 
               Spectral type  &  & F2IV + K2IV & 2 \\
        Primary mass & $m_A$ &  1.5   $M_{\sun}$  & 2 \\ 
     Secondary mass & $m_B$ &  0.8    $M_{\sun}$ & 2 \\ 
     Primary radius & $r_A$ & 2.6 $R_{\sun}$ (0.27 mas) & 2 \\
 Secondary radius & $r_B$ & 3.4 $R_{\sun}$ (0.35 mas) & 2 \\
              Reference epoch [phase 0.0] \tablenotemark{b} & $T$ & 2445766.655  & 2 \\
        Eccentricity & $e$ & 0.00     & 2 \\
         Inclination\tablenotemark{e} & $i$ & 171.1\degr   & 2 \\ 
         Semi-major axes & $a_1, a_2$ & 0.017 AU (0.38mas), 0.032 AU (0.71 mas) & 2 \\
         Component separation & $a_1 + a_2$ &  0.049 AU (1.09  mas) & 2 \\ 
            Period & $P$ & 2.613214   day & 3 \\ 
\cutinhead{Derived values}
Longitude of ascending node\tablenotemark{c} & $\Omega$ & 89\degr $\pm$ 10\degr & \\ 
           R.A. proper motion & $\mu_{\alpha} \cos \delta$ &  85.62 $\pm$  0.07  mas yr$^{-1}$  & \\ 
  Dec. proper motion & $\mu_{\delta}$ & -9.68 $\pm$  0.07  mas yr$^{-1}$ & \\ 
       Fiducial\tablenotemark{d} R.A. & $\alpha_0$ &13:34:47.75949 $\pm$  0.00008  &  \\ 
Fiducial\tablenotemark{d} Declination & $\delta_0$ & 37:10:56.7605 $\pm$  0.0006   &   
\enddata
\tablenotetext{a}{References: (1) \citet{Lestrade:1999} , (2) \citet{Eker:1987}, (3) \citet{Mayor:1987} }
\tablenotetext{b}{Secondary is in conjunction (in front) \citep{Eker:1987}}
\tablenotetext{c}{Assumes radio centroid is centered on K-subgiant (see text).}
\tablenotetext{d}{Center of mass position at fiducial epoch (JD = 2449028.000)}
\tablenotetext{e}{Inclination from \citet{Eker:1987}, but corrected for CW rotation (c.f section 3).}
 \label{table:hr5110-params}
\end{deluxetable*}

%The angular separation between the primary phase calibrator and HR5110 was 5.5\degr, compared with 3.5\degr to the secondary calibrator.

All epochs were calibrated and imaged using the NRAO Astronomical Image Processing Software package \citep[AIPS,][]{Greisen:2003}.  Standard VLBI amplitude and delay-rate  corrections were applied,  followed by the transfer of complex gain corrections from the primary calibrator (J1308+3546) to the target (HR5110) and secondary phase calibrator (J1310+3220) \citep{Diamond:1995}.  We assumed a flux density of 0.6 Jy for J1308+3546.  After  calibration, we imaged all sources using Fourier inversion and numerical deconvolution (AIPS task IMAGR using the Clark CLEAN algorithm).
We used the nodding phase-referencing scheme \citep{Lestrade:1990},  cycling at three minute intervals between J1308+3546  and HR5110.  We observed J1310+3220 once every 15 minutes to check the  stability of the phase referencing scheme.  

The nodding phase-referencing technique requires accurate  interpolation of phases from calibrator scans to  target  scans. This proved to be problematic, both because of the the relatively large angular separation\footnote{Previous phase-referenced VLBI observations of HR5110 \citep{Lestrade:1999,Ransom:2003} used several angularly closer phase calibrator sources, but at lower frequencies (5, 8~GHz).  To check their suitability at 15~GHz, in April 2012 we made 'snapshot' VLBA observations of several angularly nearby compact sources. We found that the previously-used  sources were heavily resolved at 15~GHz and therefore unusable, and that the closest suitable compact source was J1308+3546. } between phase calibrator and HR5110 (5.5\degr), but also because all epochs were scheduled during summer months, which often results in large short-timescale phase fluctuations caused from wet-component tropospheric variations \citep[]{Pradel:2006}.  Although we able to successfully transfer  phase solutions from calibrator to target scans, it required careful inspection and flagging of excessive phase jumps (AIPS task SNFLG). We flagged all time-baseline visibilities for which the phase difference between adjacent 30 second calibrator phase solutions exceeded 60$\degr$, indicating a possible phase wrap ambiguity over the three-minute nodding cycle. This resulted in more than half the visibilities in each ten-hour observing epoch being flagged.

In spite of this significant data loss, the positional accuracy of the radio centroid positions was only modestly degraded. To establish the positional accuracy at each epoch, we measured variations in the measured position of the secondary phase calibrator (J1310+3220, angular separation 3.3\degr\ from J1308+3546) compared with its ICRF position. At all epochs but epoch A (2012.495), the measured position was within $\pm$0.1 mas of its ICRF position. This is consistent with the extrapolated uncertainty estimate of \citet{Pradel:2006}, who studied the astrometric accuracy of VLBA phase-referenced observations as a function of source-calibrator angular separation and declination for a range of tropospheric conditions. 

The expected astrometric accuracy of phase-referencing can be estimated by linearly extrapolating the RMS position uncertainty trend given in Figure~2 of \citet{Pradel:2006}. 
For a source at declination  $\delta =+25$\degr, and source-phase reference angular separation $\Delta\theta$ degrees, at an observing frequency 15~GHz and with a significant wet tropospheric component this can be written as
$$
\sigma(\Delta\theta)\sim 30\ \mu\rm{as} \times \Delta\theta 
$$ 
where $\sigma$  is the positional uncertainty in micro-arcsecond for  VLBA observations. 
Hence, the expected total RMS positional uncertainty at 3.3\degr separation (primary-secondary calibrator) is 0.1 mas, in excellent agreement with our observations, and 0.17 mas at 5.5\degr separation (primary-target separation).   We have used this estimate for the formal position uncertainties at epochs B-F. For epoch A, the peak-to-peak position jitter of the secondary calibrator was 0.2 mas, possibly resulting from an excessively wet troposphere at several telescopes. Scaling this uncertainty to 5.5\degr separation resulted in an estimated  uncertainty of 0.33 mas for the position of HR5110 at Epoch A.

The radio centroid could move by up to 0.7 mas over the course of each ten hour observing epoch, if the source were located at the KIV secondary.  This position shift would significantly smear out the source brightness in maps made using the entire 10-hour observation.  The motion can be corrected by adjusting the map phase center (using AIPS task CLCOR) with a model of the source location within the orbit versus time. This is problematic since the source position is not known a priori! In our case, this correction was unnecessary, since the flagging described above resulted in using only $\sim3$ hr contiguous unflaggged time intervals at each epoch. These were typically near the end of each 10-hour observation. Because the source motion on this timescale is less than a resolving beam, there was no significant  smearing effect.  

In addition to the VLBI observations in this paper, we also calibrated and imaged five unpublished phase-referenced observations of HR5110 from the public VLBA archive.  Although these observations lack the spatial resolution and temporal sampling cadence to determine the position of the radio source within the binary orbit, the measured positions, along with the published position of \citet{Lestrade:1999}, allowed us to make a more precise determination of the proper motion  of HR5110. 

 \begin{deluxetable*}{cccccc}[h!]
%\tabletypesize{\small}
\tablecaption{J2000 Heliocentric Positions}
%\tablewidth{6.5 in}
\tablehead
{
\colhead{JD} & 
\colhead{Year} & 
\colhead{Rt. Asc. (13:34:+)} & 
\colhead{Decl. (37:10:+)}  &
\colhead{Phase Reference.\tablenotemark{a}}  &
\colhead{Ref.\tablenotemark{b}}
}
\startdata
\cutinhead{Archival and previously published}
2449028.000	&   1993.108	&     47.75947 $\pm$ 0.00003	&   56.7601 $\pm$ 0.00070 &	OP326	&	 1 \\
2451270.833	&   1999.250	&     47.80353 $\pm$ 0.00020	&   56.6997 $\pm$ 0.00200 & 	J1340+3754	&	2	 \\
2451276.833	&   1999.266	&     47.80374 $\pm$ 0.00020	&   56.7003 $\pm$ 0.00200 & 	J1340+3754	&	2	 \\
2451631.831	&   2000.239	&     47.81060 $\pm$ 0.00004	&   56.6915 $\pm$ 0.00060 & 	J1317+3425	&	3	 \\
2454560.812	&   2008.258	&     47.86794 $\pm$ 0.00003	&   56.6129 $\pm$ 0.00030 & 	J1340+3754	&	4	\\ 
2454562.813 	&   2008.264	&     47.86799 $\pm$ 0.00003   &   56.6147 $\pm$ 0.00020 & 	J1340+3754	&	4	 \\ 
\cutinhead{This paper}
2456108.458	&   2012.495	&     47.89844 $\pm$ 0.00002	&   56.5721 $\pm$ 0.00034 & 	J1308+3546	&	5	  \\
2456118.438	&   2012.523	&     47.89852 $\pm$ 0.00003	&   56.5722 $\pm$ 0.00017 &	J1308+3546	&	5  \\
2456119.427   	&   2012.525	&     47.89865 $\pm$ 0.00001	&   56.5725 $\pm$ 0.00017 & 	J1308+3546	&	5	 \\
2456122.417	&   2012.534	&     47.89867 $\pm$ 0.00001	&   56.5726 $\pm$ 0.00017 & 	J1308+3546	&	5	 \\
2456133.385	&   2012.564	&     47.89883 $\pm$ 0.00001   &   56.5728 $\pm$ 0.00017 & 	J1308+3546	&	5	 \\
2456134.396	&   2012.566	&     47.89888 $\pm$ 0.00001   &   56.5715 $\pm$ 0.00017& 	J1308+3546	&	5	
\enddata
\label{table:vlbi-obs}
\tablenotetext{a}{Positions in reference to the calibrators: OP326 (RA=13:17:36.49418 Dec=34:25:15.9326), J1340+3754 (RA=13:40:22.9518 Dec=37:54:43.835), J1317+3425 (RA=13:17:36.4942 Dec=34:25:15.933), J1308+3546 (RA = 13:08:23.70911 Dec = 35:46:37.164)}
\tablenotetext{b}{References: 1. \citet{Lestrade:1999}, 2. VLBA archive code BG087, 3. VLBA archive code BB120, 4. VLBA archive S9644, 5. This paper.}
%\tablecomments{List of VLBI observations of HR5110 with heliocentric positions.}
\end{deluxetable*}

\begin{deluxetable*}{ccccccc}[j!]
\tablecaption{Observed Fluxes and  Gaussian model parameters}
%\tablewidth{5. in}
%\tablecolumns{6}
\tablehead
{
\colhead{Epoch} &
\colhead{Code} &
\colhead{$\Phi$\tablenotemark{a}} &
\colhead{Stokes I (mJy)} & 
\colhead{Stokes V/I} &
\colhead{Diameter (mas)\tablenotemark{b}} &
\colhead{log($T_B$)}
}
\startdata
\vspace{1mm}
% Epoch        Obscode         phase          I    	FWHM	TB
2012.495 	& A   &   0.54	&    	2.6 $\pm$ 0.6	&  $<$0.28  		&		$<$1.04  			&	$>$7.1\\ \vspace{1mm}
2012.523	& B   &   0.37	&     	7.7 $\pm$ 0.8    	&  $<$ 0.08 	&		0.76 $\pm$ 0.20		&	7.8 $\pm$ 0.1 \\ \vspace{1mm}
2012.525 	& C   &   0.73	&  	76.5 $\pm$ 5.0	&  +0.02 $\pm$ 0.01 	&	0.86 $\pm$ 0.05	&	8.7 $\pm$ 0.1\\ \vspace{1mm}
2012.534	&  D  &   0.88	&   	91.6 $\pm$ 0.4	& +0.04 $\pm$ 0.01	&	0.36 $\pm$ 0.05		&	9.6 $\pm $ 0.1 \\ \vspace{1mm}
2012.564	&  E  &    0.11	&   	12.2 $\pm$ 0.9 	&+0.18 $\pm$ 0.01   	&            0.94 $\pm$ 0.20		&	7.9 $\pm$ 0.1 \\  \vspace{1mm}
2012.566	&  F  &    0.47	&  	10.3 $\pm $1.6	&	$<0.12$       	&  1.08 $\pm$ 0.20		&	7.7 $\pm$ 0.1
\enddata
\tablenotetext{a}{Orbital phase computed using  the period in Table~\ref{table:hr5110-params} and adopting phase 0 at secondary conjunction \citep{Eker:1987}. }
\tablenotetext{b}{Sizes are the full-width at half maximum geometric means of  elliptical single component Gaussian fits.}
\label{table:gaussian-fits}
\end{deluxetable*}

 \begin{figure*}[h!]
\centering
\includegraphics[width =5.in]{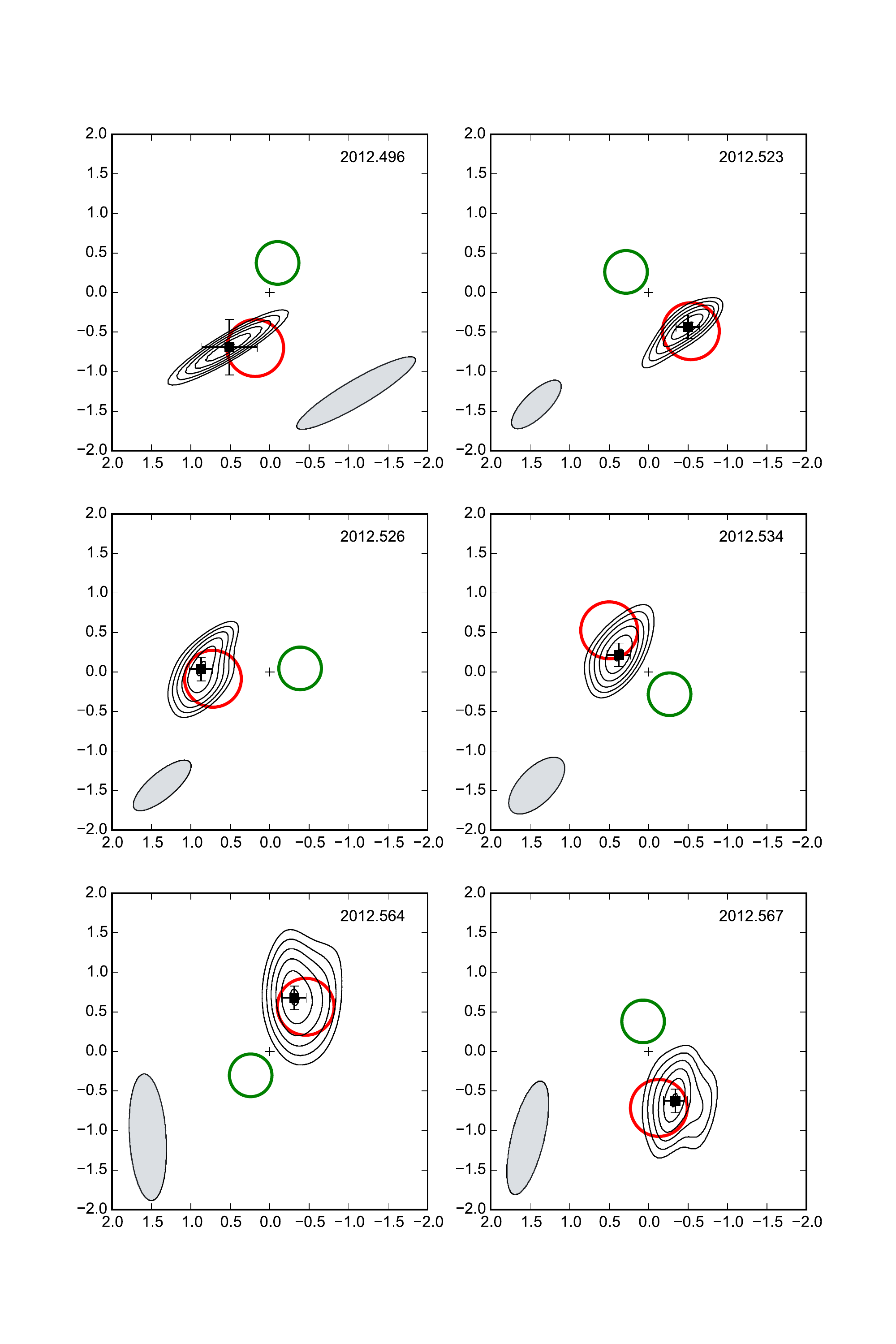}
\caption{ Radio contour maps of HR5110 at 15 GHz overlaid on the binary system in the center of mass frame of reference. The contour intervals are: 50, 60, 70, 80, 90 percent of the maximum radio intensity at each epoch.  The axes label ticks are in mas relative to the binary center of mass.
The superposed star positions (green circle= FIV primary, red circle = KIV secondary) are consistent with a least-squares minimization solution for a fixed location of the radio centroids (at all six epochs) in the co-rotating frame of the binary system (see section 3).  Since the resulting center of mass radial displacement was close to the K-star displacement, we identify the radio source location with the K secondary.   
The grey ellipse is the restoring beam at each epoch.}
\label{fig:BM359 contour maps}
\end{figure*}

\section{Results}
We determined new proper motions  for HR5110 (Table~\ref{table:hr5110-params}) by combining our 6-epoch position determinations with the previously published position of \citet{Lestrade:1999} and positions that we determined from archival VLBA observations, as listed in Table~\ref{table:vlbi-obs}. Our calculated  right ascension proper motion is in good agreement with the \citet{Lestrade:1999} value, while the declination proper motion is three standard deviations from their value.

Radio contour maps of HR5110 at all six epochs are shown in Figure \ref{fig:BM359 contour maps}. Previous VLBI observations of HR5110 found a core-halo source morphology  at 5~GHz \citep{Mutel:1985} and 8~GHz \citep{Ransom:2003}. The 15~GHz maps are somewhat amorphous and do not manifest a clear core-halo structure even if the long baselines to the more sensitive Effelsberg telescope are excluded. However, we can characterize the spatial scale sizes and fluxes by fitting the images with a one component Gaussian. Fluxes vary significantly and are likely underestimates of the true flux due to the rapidly varying wet troposphere introducing significant phase errors \citep{Marti-Vidal:2010}.  All epochs were fit using a single component Gaussian and the resulting 
Gaussian model parameters are given in Table \ref{table:gaussian-fits}.  

\begin{figure}
\centering
\includegraphics[width=3.5 in]{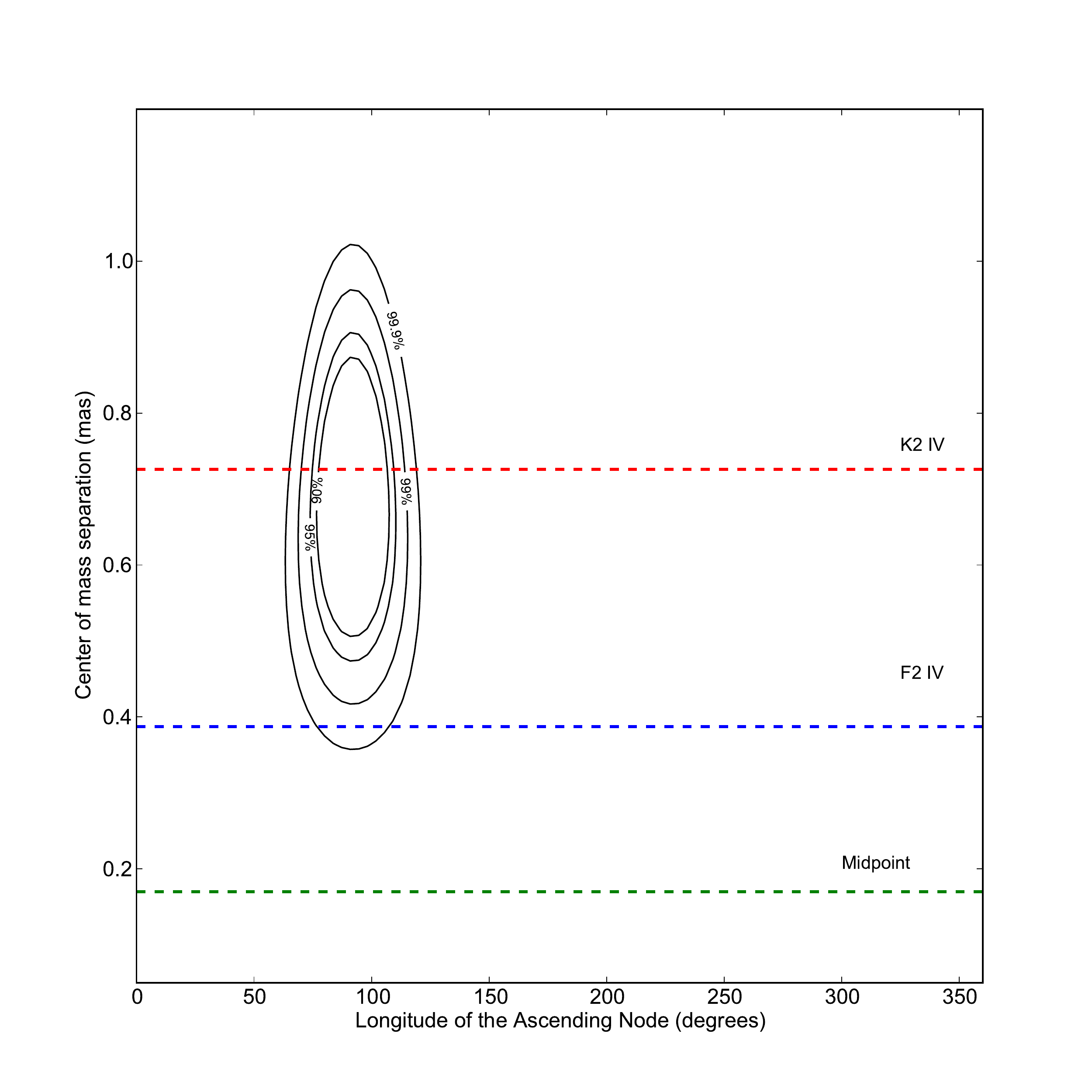}
\caption{Confidence intervals of best-fit solution (center of mass displacement, ascending node angle) for stationary location of radio emission in the co-rotating frame. The dotted lines indicate the radial offsets from the center of mass for the K2 IV star (red), F2 IV star (blue) and the midpoint between the two (green).  Contours represent confidence intervals centered on the best-fit position ($r = 0.70$ mas, $\Omega = 89\degr$).   }  
\label{fig:Orbital Parameter Minimization}
\end{figure}

We used our measured proper motions, along with the parallax determined by  \citet{Lestrade:1999}, to convert the observed radio centroids  to displacements relative  to the binary center of mass. These displacements were  used to solve for the co-rotating location of the radio source with respect to the binary's center of mass. To do this, we assumed that the radio emission at all epochs is located at a fixed point in the co-rotating frame. We solved for this location by finding the point in the co-rotating frame that minimizes the sum  of the squared distances (weighted by position uncertainty) from the center of mass to the radio centroids.  Figure \ref{fig:Orbital Parameter Minimization} shows the $\chi ^2$ minimization surface as a function of the radial displacement from the center of mass and an azimuthal angle measured north through east in the sky plane.   Table~\ref{table:position-offset} lists the positional offset at each epoch between the measured radio centroid and the best-fit fixed point in the co-rotating frame.

\begin{figure}
\centering
\includegraphics[width=3.5 in]{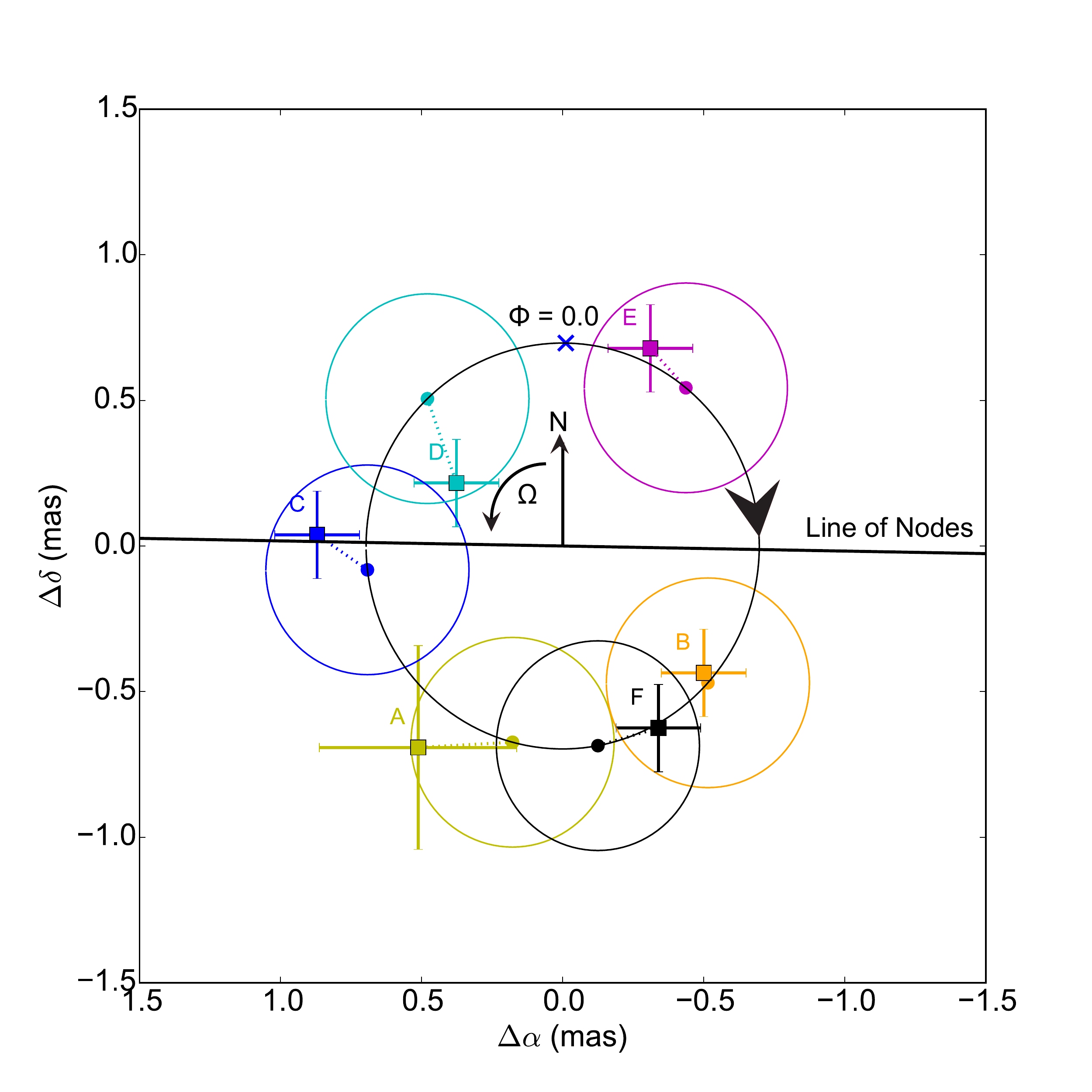}
\caption{HR5110 radio centroid locations at six epochs in the co-moving center of mass system. Each epoch is color coded:  A (yellow),  B (orange), C (blue), D (cyan), E (magenta), F (black).  The large black circle is the orbit of the cooler KIV secondary star.  Smaller colored circles show the secondary's size and location at the time of the corresponding observation. The dotted lines indicate the distance between the radio centroid and the KIV center at each epoch.}
\label{fig:Center of Mass Positions}
\end{figure}

\begin{figure}
\centering
\includegraphics[width=3.5 in]{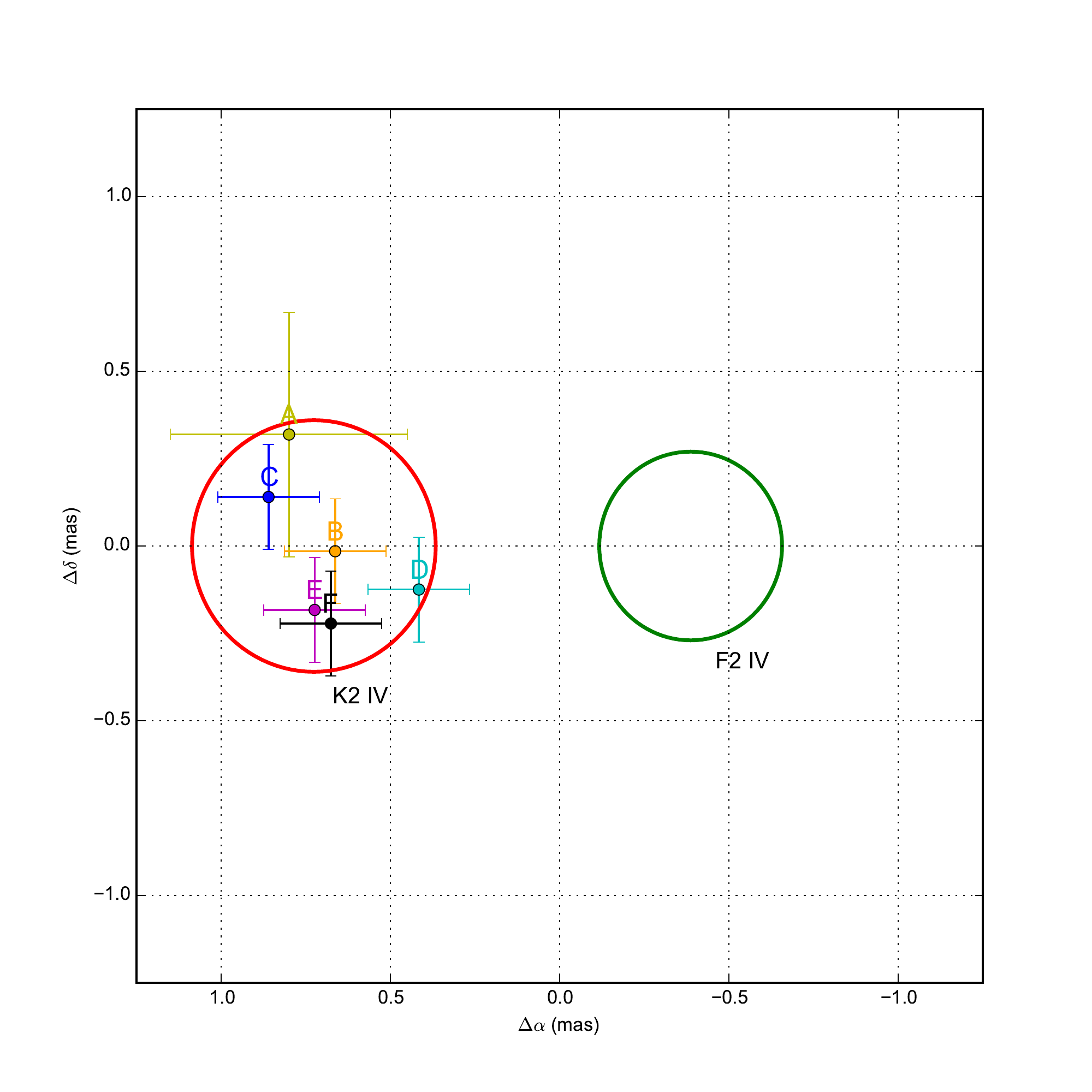}
\caption{Radio centroid positions overlaid on the binary system in the co-rotating frame.  The orientation of the stars is based on the close association of the best fit radial offset with the K2 IV semi-major axis.  Each epoch is color coded:  A (yellow),  B (orange), C (blue), D (cyan), E (magenta), F (black).  The two circles are the K2 IV star (red) and the F2 IV star (green).  }
\label{fig:Corotating Positions}
\end{figure}

The best fit radial displacement is nearly the same as the semi-major axis of the K2 IV secondary, rather than the F2 IV primary or an interaction region midway between the stars.  Since the binary's orientation on the sky is not known,  the radio centroid azimuthal angles cannot be directly compared with the location of the component stars. However, given the close agreement between the model's radial displacement and the K star's semi-major axis, we  make the plausible assumption that the radio centroids are  located on or near the K star rather than a random co-moving location on the K star's orbit.  Figure~\ref{fig:Center of Mass Positions} shows the radio centroids and star positions in  co-moving sky plane coordinates, while Figure \ref{fig:Corotating Positions} is in the co-rotating frame, assuming the radio centroid-K star identification described above.

\begin{deluxetable}{ccc}
\tablecaption{Offsets from best-fit model position}
\tablewidth{2in}
\tablehead
{
\colhead{Epoch} & 
\colhead{$\Delta$$\alpha$  ($\sigma$)\tablenotemark{a}} & 
\colhead{$\Delta$$\delta$  ($\sigma$)\tablenotemark{a}}  
}
 
\
\startdata
1993.108	&     0.0 (0.0)	&  -0.9 (1.3)	 \\
1999.250	&     1.1 (0.6)	&   -1.0 (0.5)	 \\
1999.266	&     1.4 (0.7)	&   0.1 (0.1) 	 \\
2000.239	&     0.7 (1.4)	&   0.8 (1.4) 	 \\
2008.258	&     -0.3 (1.0)	&   -0.7 (2.5) 	 \\
2008.264	&     -0.5 (1.7)   &   0.3 (1.1)  	 \\
2012.495 &    0.6 (3.9)	&     -0.1 (0.6) 	 \\ 
2012.523	&     -0.2 (0.6)	&    0.1 (0.3)	 \\ 
2012.525 &        0.2 (1.5)	&    0.0 (0.0)	 \\ 
2012.534	&        -0.1  (0.4)	&    -0.4 (2.6)	 \\ 
2012.564	&      0.0 (0.1)	&     0.1 (0.8)	 \\  
2012.566	&      0.0 (0.2)	&     0.0 (0.1)
\enddata
\label{table:position-offset}
\tablenotetext{a}{Observed - model in mas (std. dev.)}
%\tablenotetext{d}{J2000 Declination, 37:10 +}
%\tablenotetext{e}{References: 1. \citet{Lestrade:1999}, 2. VLBA archive code BG087, 3. VLBA archive code BB120, 4. VLBA archive S9644, 5. This paper.}
%\tablecomments{List of VLBI observations of HR5110 with heliocentric positions.}
\end{deluxetable}

% Center of mass figure

The identification of the radio centroid with the K IV secondary allows us to determine new values for two orbital elements of the binary system. First, since the observed radio centroid motion in the center-of-mass system is clockwise on the sky (Figure~\ref{fig:Center of Mass Positions}, black arrow on secondary orbit), the inclination angle is in the second quadrant \citep{Heintz:1978}.  Hence $i$ = 171.1\degr\ rather than 8.9\degr\  as given by \citet{Eker:1987}. Second, the longitude of the ascending node (labeled $\Omega$ in Figure~\ref{fig:Center of Mass Positions}), which was previously unknown, was determined by using the epoch of secondary conjunction \citep[][ phase 0]{Eker:1987} to determine the interpolated radio position at phase 0.75, i.e. the time of transit through the ascending node. We find $\Omega=89\degr\pm10\degr$ where the uncertainty corresponds to the scatter in   co-rotating radio centroid positions (Figure~\ref{fig:Corotating Positions}) translated to orbital phase.

\section{Discussion}
\subsection{ Active Region Location}

The principal astrometric result of this study is that the radio centroids of HR 5110 are located within one stellar radius of the co-rotating K~IV secondary center at all epochs. The identification of the radio emission with the secondary component in close binaries has  been seen previously in several other such systems viz., Algol \citep{Lestrade:1993,Peterson:2010}, UX Ari \citep{Peterson:2011}, and HR8703 \citep{Ransom:2012}.  Similarly, \citet{Little-Marenin:1986} found that UV emission lines associated with an active chromosphere and transition region are located on the KIV secondary.  By contrast, \cite{Graffagnino:1995} modeled  a very large X-ray flare from HR5110 and found that the source size was comparable with the binary separation. They suggested that the interaction region between the stars could be the origin of the emission, e.g., the interbinary magnetic region posited for the  binary CZ CVn \citep{Strassmeier:2011}.  In addition, during an extreme radio flare Algol's emission was clearly displaced towards the hotter component \citep{Peterson:2010}.  

This apparent disagreement concerning the active region location poses an interesting conundrum: Are these systems driven by magnetic activity on the cooler star or between the stars?  It is possible that the size and location of the active region depends on the intensity and duration of the flare.  For all our  VLBI epochs, the observed radio flux density of HR5110 was significantly less than the highest recorded flux outburst \citep[$S>400$ mJy,][]{Feldman:1980}.  However, the large HR5110 X-ray event and the large radio loop observed on Algol represent extraordinarily large outbursts.  It is possible that  moderate flares are located on the cooler secondary but that  emission regions responsible for the largest flares can extend into, or arise from an interbinary  region.

Previously the radio emission was thought to be tied to a magnetically active polar spot \citep{Elias:1995}, but the VLBI determined radio positions are not tightly grouped on a scale smaller than a stellar radius.  Polar spots in binary systems are thought to be long lived, up to several years \citep{Strassmeier:2009}.  By contrast the scatter of radio positions seen in Figure \ref{fig:Corotating Positions} occurs over a time period only slightly more than a month.  In HR5110 the orbital plane is inclined nearly perpendicular to the line of sight.  If the magnetic pole of the cooler star is roughly aligned with its rotation axis then a putative polar spot would be nearly stationary from the observer's position.  The observed scatter in the HR5110 radio centroids is inconsistent with emission from a stationary location associated with a polar spot.  This result for HR5110 is very similar to HR8703 \citep{Ransom:2012} where the radio centroids at multiple, well-separated epochs are scattered over the entire stellar disk.

\subsection{Physical Properties of Emission Region: Gyrosynchrotron Model }

% GS model parameter surface
\begin{figure*}
\centering
\hspace*{0in}
\includegraphics[width=6.5in]{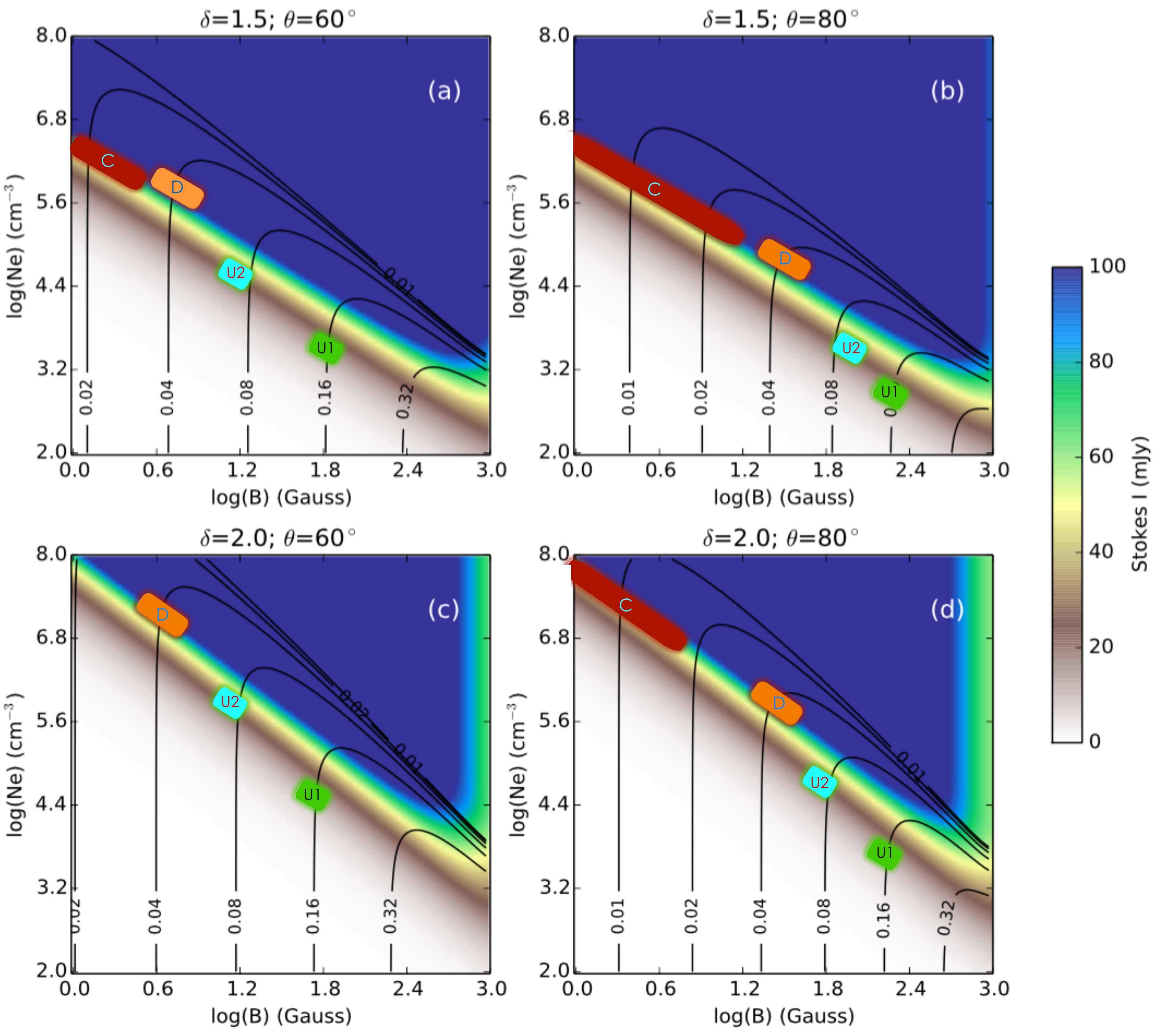}
\caption{Comparison between observed and model  15~GHz flux density (colored surface) and fractional circular polarization (CP, black contours) for HR5110. Model fluxes were calculated using a uniform box of size 5.8~$R_{\sun} $ (0.6 mas) at 45 pc emitting gyrosynchrotron radiation for two different magnetic field orientations  and electron power-law indices: (a) $\theta$~=~60\degr, $\delta$~=~1.5; (b) $\theta$~=~80\degr, $\delta$~=~ 1.5; (c) $\theta$~=~60\degr, $\delta$~=~2.0; (d) $\theta$~=~80\degr, $\delta$~=~2.0. Also shown are rectangular boxes delineating observed flux densities and fraction CP (with uncertainties) at epochs  2012.525 (C, red)  2012.534 (D, orange), as well as 15~GHz fluxes and CP reported by \citet{Umana:1993} at epochs 1989.134 (U1, green) and 1989.178 (U2, cyan). }
\label{fig:GS parameters}
\end{figure*}

To characterize the plasma conditions responsible for the observed radio emission, we constructed a simplified coronal model consisting of constant density, mildly-relativistic power-law electrons in a uniform magnetic field emitting gyrosynchrotron emission.  The emitting volume is a cube whose  size is the average observed VLBI angular size ($0.6 {~\rm mas} = 5.7R_{\sun}$).  This is clearly an oversimplified model, but allows for order-of-magnitude estimates of the electron density and magnetic field strength without making unsubstantiated assumptions concerning the magnetic field topology or energetic electron population parameters.  

The model solves for the emergent Stokes I and V flux density at the distance to HR5110 (45 pc.). We used the expressions of \citet{Robinson:1984} for the emission  and absorption coefficients of gyrosynchrotron emission from mildly relativistic electrons with a power-law electron energy distribution  given by,
$$
N_e(E) = K E^{-\delta},
$$
where $K =N_0(\delta-1){E_0}^{\delta-1}$ and  is $\delta$ the energy index, and with a low-energy cutoff $E_0$ = 10~keV.
The emergent flux density was calculated by integrating the radiative transfer equation, summing the contributions along uniformly spaced lines of sight through the  cube  
We varied  the strength and orientation of the magnetic field as well as the density and energy index of the relativistic electron population  to best-fit the measured 15~GHz integrated flux density and fractional circular polarization at each epoch.

Figure \ref{fig:GS parameters} shows a surface plot of Stokes I flux density as a function of electron density and magnetic field strength for four pairs of energy index and magnetic field orientation angle $\theta$ measured with respect to the observer's line of sight. Multi-frequency observations of HR5110 were obtained by \citet{Umana:1993} and \citet{White:1990a}. Both reported relatively flat spectral indices ($\alpha \leq$ -0.6) at 15~GHz.  For a synchrotron process, the spectral index is simply related to the power-law electron energy index $\delta = 1-2\alpha\leq 2.2$, so we chose representative values  1.5 and 2.0. The magnetic field orientation is unconstrained by prior observations, but values smaller then 60\degr\ resulted in model circular polarization fractions that exceeded our observed values, so we show only larger angles. 

For a given Stokes I flux density the solution is highly degenerate, with constant flux density values occurring along contours of nearly constant slope in log($N_e$)-log($B$) coordinates. These lines have a slope that depends on primarily on the energy index, and can be approximated by
$$\frac{\log(N_e)}{\log(B)}\sim-0.8\delta$$
However, a measurement of circular polarization (CP) can break this degeneracy,  as shown by the black contours, which are labeled by fractional CP. 
At epochs C and D, the signal-to-noise ratio was sufficient to measure significant non-zero circularly polarized emission (Table \ref{table:gaussian-fits}), so we used these epochs to constrain the model.  Observed flux densities and fractional circular polarization are shown as overlaid rectangles (red: epoch C, orange: epoch D), where the widths represent the uncertainty in each value.  For comparison, we also  show 15~GHz flux densities and CP  reported by \citet{Umana:1993} at epochs 1989.134 (green, label U1) and 1989.178 (cyan, label U2). 

Comparing the model with observed fluxes and CP, we find that although there is still a significant parameter degeneracy, the present observations favor electron densities in the range  $n_e\sim10^5$ cm$^{-3}$ and magnetic field strengths B$\sim 10$ Gauss, but with a large allowable range (approximately $\pm\ 1$ dex). By comparison, the \citet{Umana:1993} observations favor somewhat lower electron densities and higher magnetic field strengths.  Since the source is highly variable, these differences may reflect different physical conditions in the emission region at different epochs.

\section{Summary}

We observed the close active binary HR5110 with  a global VLBI array at 15.4 GHz at six epochs distributed uniformly over the binary's orbit phase. The primary goal of these observations was to determine the location of the radio source within the binary system. To do this, we used a phase-referencing nodding scheme, rapidly switching from the target star to an angularly nearby extragalactic radio source. 

After correcting for parallax and proper motion, we determined the locations of the radio centroids within the co-rotating binary system with sub-millarcsecond accuracy.
The  positions were scattered around a fixed location [within measurement uncertainty] in the co-rotating frame of the binary. The radial offsets from the center of mass were nearly the same as the  semi-major axis of the KIV secondary, strongly suggesting the KIV secondary as the source of the radio emission.  This identification of the radio source with the cooler, more evolved star has been previous reported for several other close binary systems, including Algol \citep{Lestrade:1999,Peterson:2010}, UX Ari \citep{Peterson:2011}, and IM Peg \citep{Ransom:2012}.  

The clockwise radio centroid motion indicates an orbital inclination angle of 171.1$\degr$ rather than $8.9\degr$ as given by \citet{Eker:1987}. Additionally, the longitude of the ascending node ($\Omega = 89\degr$) was determined using the identification of the radio centroids with the K~IV secondary and the previously determined time of secondary conjunction \citep{Eker:1987}.

We used the observed properties of the radio emission (flux density, angular size, circular polarization) to constrain  plasma parameters in the source region. The model used a simplified geometry consisting of power-law relativistic electrons in a uniform magnetic field radiating gyro-synchrotron emission. The  derived values of magnetic field strength and relativistic electron density are degenerate, but the degeneracy is broken when there is a measured fractional circular polarization.  For two such epochs, we find a mean magnetic field strength of order 10 Gauss and electron densities that vary from $10^4 $\ cm$^{-3} - 10^6$ cm$^{-3}$ depending on the electron power-law index and magnetic field orientation.

\vspace{0.2in}
We are grateful for financial support  from the National Science Foundation through research grant AST-0908941. This research has made use of the SIMBAD database, operated at CDS, Strasbourg, France.

\bibliography{rlm-library}

\end{document}